# ANALYSIS OF MINNESOTA COLON AND RECTUM CANCER POINT PATTERNS WITH SPATIAL AND NONSPATIAL COVARIATE INFORMATION[1]

By Shengde Liang, Bradley P. Carlin and Alan E. Gelfand

*University of Minnesota, University of Minnesota and Duke University*

Colon and rectum cancer share many risk factors, and are often tabulated together as "colorectal cancer" in published summaries. However, recent work indicating that exercise, diet, and family history may have differential impacts on the two cancers encourages analyzing them separately, so that corresponding public health interventions can be more efficiently targeted. We analyze colon and rectum cancer data from the Minnesota Cancer Surveillance System from 1998–2002 over the 16-county Twin Cities (Minneapolis–St. Paul) metro and exurban area. The data consist of two marked point patterns, meaning that any statistical model must account for randomness in the observed locations, and expected positive association between the two cancer patterns. Our model extends marked spatial point pattern analysis in the context of a log Gaussian Cox process to accommodate spatially referenced covariates (local poverty rate and location within the metro area), individual-level risk factors (patient age and cancer stage), and related interactions. We obtain smoothed maps of marginal log-relative intensity surfaces for colon and rectum cancer, and uncover significant age and stage differences between the two groups. This encourages more aggressive colon cancer screening in the inner Twin Cities and their southern and western exurbs, where our model indicates higher colon cancer relative intensity.

**1. Introduction.**

1.1. *Etiologies of colon and rectum cancer.* Traditionally, public health agencies have reported colon and rectum cancers together under the title "colorectal cancer." Since the turn of the last century, however, an active debate has emerged regarding whether these two cancers really have sufficiently similar etiologies to be aggregated in this way. Some experts have

Received May 2008; revised February 2009.
[1]Supported in part by NIH Grant 1–R01–CA95955–01.

argued that the cancers should be reported and monitored separately, so that public health interventions can be more sensibly and efficiently targeted.

A variety of epidemiological studies have indicated variables that may have a differential impact on colon and rectum cancer. These variables fall under three broad categories. The first is exercise. A very recent study by the Physical Activity Guidelines Advisory Committee (2008) identified 23 publications on this general topic, arising from 12 prospective cohort studies and 8 case-control studies. These studies show a consistent inverse relation between physical activity and colon cancer risk, with this relation being statistically significant for at least one physical activity domain and one sex in 9 of the 12 cohort studies and 5 of the 8 case-control studies. More specifically, the median relative risk (RR) comparing most- versus least-active subjects was 0.7 over all the studies. The advisory committee stated that this overall finding was unlikely to be the result of confounding, since the studies for the most part included relevant covariates, such as body mass index (BMI), smoking, alcohol, diet, screening, menopausal status, and family history of colon cancer. By contrast, the committee found the studies to indicate no apparent relationship between physical activity and rectal cancer risk. Specifically, more than half the studies showed no statistically significant associations, and the median RR over all the studies was 1.0. The fact that moderate-to-vigorous physical activity (say, 30 to 60 minutes per day) may be protective against colon cancer but not rectal cancer suggests that these two cancers should be treated separately in cancer registry reporting and subsequent statistical modeling.

A second broad area that may have a differential impact on the two cancers is diet. Diet has long been suspected as an etiological factor for colorectal cancer; however, studies of individual foods and nutrients have often provided inconsistent results, perhaps due to low statistical power. Flood et al. (2008) address this problem using factor analysis to group dietary variables into three broad groups, and go on to conclude that lower consumption of meat and potatoes, and higher consumption of fruit, vegetables, and fat-reduced foods, are associated with reduced colorectal cancer risk. Wei et al. (2003) use data from two prospective cohort studies (87,733 women from the Nurses' Health Study and 46,632 men from Health Professionals Follow-Up Study) to investigate the effect of dietary variables on colon and rectum cancer separately. In the combined cohort, a variety of variables emerge as significant predictors of elevated colon cancer risk, including intake of beef, pork or lamb as a main dish, intake of processed meat, and alcohol consumption. However, none of these variables emerge as predictors of rectal cancer. Using data from the Iowa Women's Health Study, Folsom and Hong (2005) showed that magnesium and calcium intake were independently associated with significantly lower colon cancer risk, but not rectum cancer risk. The relative risk estimates changed little across baseline subgroups, such as women who did or did not use hormone replacement therapy, or were or were not diabetic. Using data from a different study, Flood et al. (2005) conclude that the protective effect of calcium is present regardless of whether the calcium arises naturally in food, or is delivered through dietary supplements. Finally, Pedersen, Johansen and Gronbaek (2003) observed a dose-response relationship between alcohol and rectal cancer in a Danish cohort of 15,491 men and 13,641 women who did not include wine in their alcohol intake. However, no association between alcohol and colon cancer was found.

The third broad area where the etiologies of colon and rectal cancer appear to differ is family history. For those who reported a family history of colon or rectal cancer, Fuchs et al. (1994) obtained a RR of 1.99 for colon cancer but just 0.86 for rectal cancer, a statistically significant difference based on a simple chi-square test with one degree of freedom. Wei et al. (2003) drew the same conclusion, but using a different dataset and a stepwise polytomous logistic regression procedure.

1.2. *MCSS data and problem description.* Our specific problem of interest involves the comparison of the spatial distributions of colon and rectum cancer patients in the state of Minnesota. These data are collected by the Minnesota Cancer Surveillance System (MCSS), a program sponsored by the Minnesota Department of Health. The MCSS includes the residential address of essentially every person diagnosed with cancer in Minnesota. Here we consider the subset of patients diagnosed during the period 1998–2002 (an interval chosen partly for its centering around a U.S. Census year, 2000). Figure 1 shows the 7 counties comprising the Twin Cities metro area as those encircled by the dark boundary; also shown are 9 adjacent, exurban counties. Within these 16 counties, we have 6544 individuals for analysis. Figure 1 plots the approximate locations of the cancers after the addition of a random "jitter" to protect patient confidentiality (explaining why some of the cases appear to lie outside of the spatial domain). The physiological adjacency of the colon and the rectum suggests positive dependence in these point patterns; persons with rectum cancer beyond stage 1 (i.e., regional or distant) are at risk for colon cancer due to metastasis. Moreover, the two cancers likely share unmodeled spatially-varying risk factors (such as local health care quality or availability), also implying positive dependence. This may help health care providers or public health policy makers to identify regions of excessive risk requiring intervention (say, a direct mail campaign encouraging more aggressive screening) or other weak links in the health care system.

The causes of colon and rectum cancer are unknown. Age is the primary risk factor, with disease incidence increasing significantly after the age of 50.

As already mentioned, family medical history may also be helpful in predicting colon and rectum cancer risk, along with several lifestyle factors such as alcohol use, smoking, diet, and exercise. Unfortunately we do not have access to this information for individuals in the MCSS, but the lifestyle factors could reasonably be expected to cluster spatially due to corresponding sociodemographic clustering. We also have census tract-level poverty rates, which should be correlated with these risk factors.

A full analysis of the data in Figure 1 would account for the randomness in the observed locations, their spatial correlation, important covariates (including population density), and any other hierarchical structure in the data (such as the tendency of model residuals to cluster spatially). The output of such an analysis would include maps of the fitted adjusted log-intensity surface, point and interval estimates for important main effects and interactions (e.g., location-age), and perhaps maps of fitted spatial residual surfaces, to help identify spatial covariates still missing from the model.

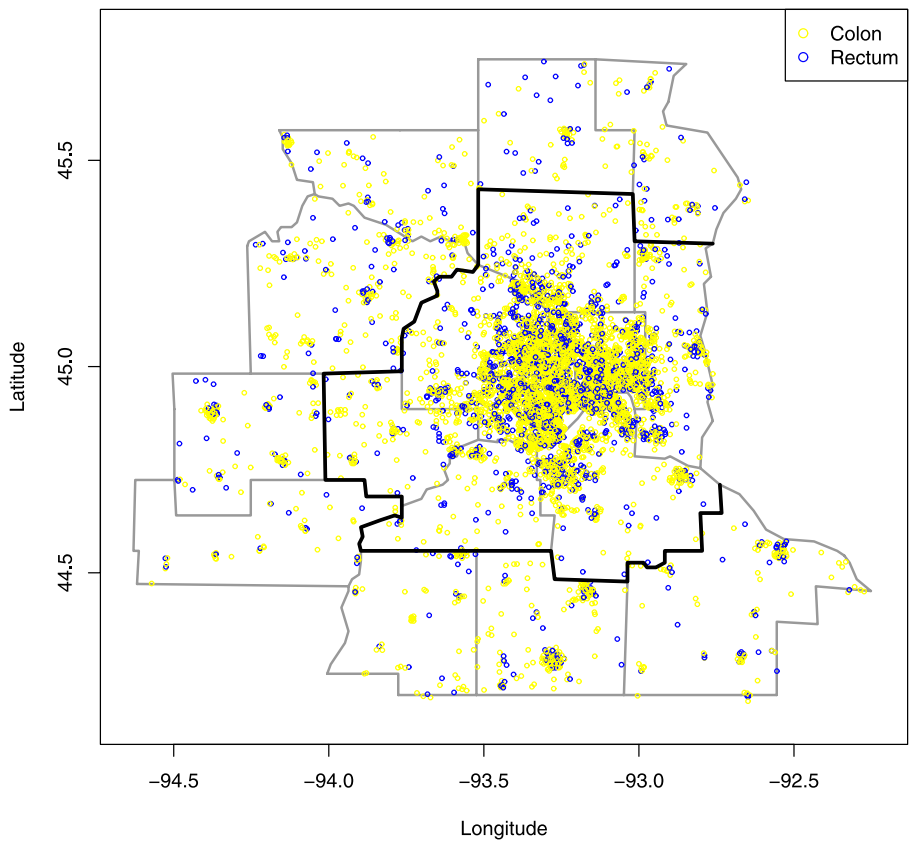

FIG. 1. *Jittered residential locations of colon (light circle) and rectum (dark circle) cancer cases, Twin Cities metro and exurban counties, 1998–2002.*

1.3. *Statistical modeling of spatial point patterns.* In spatial disease mapping settings, the primary goals are typically to investigate the connections between the disease and (possibly geographically-indexed) covariates, to characterize the spatial variation of the disease occurrence, and to identify areas having elevated disease risk. In such cases, the data are often aggregated to counts within specified areal regions (counties, zip codes, etc.). Indeed, most published statistical analyses to date of data of this type use so-called *areal* or *lattice* models; see, for example, Banerjee, Carlin and Gelfand [(2004), Chapter 3 and Section 5.4] for a review. However, if precise geocoded locations of disease cases are available, it is more appealing to study the resulting *spatial point pattern* using spatial point process modeling. However, such methods are conceptually and computationally more challenging, and are implemented in fewer widely available statistical software programs. Indeed, even when actual geocoded locations are available, a standard computational strategy is to partition the study region and model the counts in each cell of the partition as conditionally independent Poisson observations, obtaining the standard areal model but with an arbitrary partition.

Under a nonhomogeneous Poisson process, the likelihood for the *intensity surface* generating the locations given the observed locations is well known [see, e.g., Beneš et al. (2002); Diggle (2003); Møller and Waagepetersen (2003)]. We begin with this likelihood, but introduce the following features. First, we accommodate covariate information in a novel way. We envision certain covariates as conditional, that is, we seek to compare point patterns given levels of these covariates. For us, these are *cancer type* covariates which "mark" the point pattern. We view other patient level characteristics (or risk factors) such as age as nuisance variables for which we wish to adjust. We then model point patterns *jointly* over geographic space and nuisance covariate space, enabling the notions of both conditional and marginal intensity associated with geographic space. Hence, we obtain an intensity adjusted for these covariates, rather than an intensity which ignores them by not including them in the model. Moreover, we also have purely spatial covariates, some of which are available at areal unit level (say, county-level features), while others may be available at point level (say, distance from a location to the nearest cancer screening facility). Employing spatial information at both scales precludes aggregation of points to counts.

Additionally, we anticipate dependence between the intensity surfaces associated with the two cancers, since, for example, an excess of colon cancer in a portion of the study region may suggest correspondingly high levels of rectum cancer. We capture this dependence using *multivariate* process realizations for the intensities. Last, working with the above point level likelihood, as well as fairly large numbers of points (e.g., order $10^3$), necessitates approximation to implement the model-fitting.

The analysis of spatial point patterns has a reasonably long history in the literature, initially built using exploratory tools such as distance based methods yielding $F$ functions, $G$ functions, and, perhaps most commonly, Ripley's $K$ function. All are based upon assessing departure from *complete spatial randomness* (CSR), which is interpreted as a homogeneous Poisson process and for which closed forms for these functions exist. However, no likelihood is specified, and comparison between point patterns is not possible. Another more recent approach involves the use of *spatial scan* statistics, currently popular in large part due to the SatScan software of Kulldorff (2006). But again, no likelihood is specified so inference is limited to say detection of "hot spots."

To achieve the foregoing objectives, we instead adopt a model-based focus, and write the intensity of the process as $\lambda(s)$, where $s \in \mathcal{D}$ for some spatial domain $\mathcal{D}$. For a collection of observed cancer case locations $\mathbf{s}_i, i = 1, \ldots, n$, we work with the likelihood $L(\lambda(\mathbf{s}), \mathbf{s} \in D; \{\mathbf{s}_i\}_{i=1}^n)$ which takes the form $e^{-\int_D \lambda(\mathbf{s}) \, d\mathbf{s}} \prod_{i=1}^n \lambda(\mathbf{s}_i)$. Often, $\lambda(\mathbf{s})$ is specified as a parametric function, for example, using a basis representation or a tiled surface. Adding a prior distribution on these parameters, say, $\boldsymbol{\theta}$, yields a posterior distribution $p(\lambda(\mathbf{s}; \boldsymbol{\theta}) | \{\mathbf{s}_i\})$ for making inferences about the intensity surface.

For us, $\lambda(\mathbf{s})$ is thought of as a log Gaussian process (GP) realization, resulting in the familiar class of Cox processes [Møller and Waagepetersen (2004), page 57]. To specify this prior distribution, we require $\mu(\mathbf{s})$, the mean surface, along with $\sigma^2$ and $\phi$, the GP covariance parameters. Below, we express $\mu(\mathbf{s})$ in part with a form $\mathbf{z}'(\mathbf{s})\boldsymbol{\beta}$, so that the process mean can depend on spatially referenced covariates $\mathbf{z}(\mathbf{s})$.

A common class of estimation methods for inhomogeneous spatial point process models avoids full likelihood evaluations by formulating estimating equations [Waagepetersen (2007); Waagepetersen and Guan (2009)]. Guan and Loh (2007) study the distributional properties of the estimation procedure of Waagepetersen (2007), and obtain variance estimates using a thinned block bootstrap procedure. Guan (2006) developed a composite likelihood method based on the second-order intensity function of the underlying process. Diggle and Rowlingson (1994) handle bivariate (case-control) point processes via a conditional likelihood approach to convert the two spatial point process models into an easier-to-fit nonlinear binary regression model. Similarly, Guan, Waagepetersen and Beale (2008) estimate correlation functions via either a consistent nonparametric kernel smoothing estimator, or a parametric conditional likelihood estimator. In all of these approaches, inference on spatial associations and second-order variations proceeds not from the intensity surface, but from pairwise correlation functions and transforms thereof [e.g., the $g$ and $K$ functions in Waagepetersen (2007)]. As such, they do not offer direct attacks on the intensity surface estimation problem, needed for inference regarding the fitted surface itself, its rate of change at any point [as needed for spatial boundary analysis or "wombling"; see Banerjee and Gelfand (2006), and Liang, Banerjee and Carlin (2009)], or model-based comparison of the surfaces for colon and rectum cancer.

As such, we instead adopt a fully Bayesian approach that yields posterior distributions for the intensity surface, or even the spatial residual surface after adjusting for regressors that are allowed to differ for the two cancers. Due to the absence of sufficient covariate (e.g., diet) information in our dataset, we introduce spatially varying random effects which we view as surrogates for these missing covariates. Inference is exact and does not rely upon possibly inappropriate use of infill or increasing-domain asymptotics. However, our more comprehensive approach comes with a price. Specifically, note that if $\lambda(\mathbf{s})$ is modeled as a random realization of a spatial process, then the likelihood integral is stochastic, precluding explicit evaluation. Indeed, a variety of computational challenges emerge in working with the point-level likelihood in this case: the stochastic integration, the large collection of spatial locations, and a prior specification that is only available through finite dimensional distributions.

Wolpert and Ickstadt (1998) offered one of the first fully Bayesian approaches for spatially nonhomogeneous Poisson process data. Beneš et al. (2002) illustrate one possible Bayesian analysis of a log Gaussian Cox process model. While they assume $\lambda(\mathbf{s})$ constant over grid cells, they do utilize the notion of the population intensity surface, and obtain fitted disease maps under a variety of models (constant, Gaussian kernel, etc.) for this surface. However, they do not consider joint modeling of multiple disease surfaces, nor covariates that are not location-specific (e.g., the age or cancer stage of a case observed at a particular location).

In this paper we employ novel spatial point process approaches that account for both location-specific and nonlocation-specific covariates in the context of multiple dependent point processes to analyze the MCSS dataset. We begin in Section 2 with a brief review of spatial point process modeling. We then go on to present a set of multivariate spatial point process models for investigating the effect of both location-specific and nonlocation-specific covariates, as well as their interactions. Section 3 then gives the results of applying them to our MCSS dataset. Finally, Section 4 discusses our findings and offers directions for future research in this area. Computational challenges in fitting our models are addressed in the supplemental article [Liang, Carlin and Gelfand (2009)].

## 2. Hierarchical modeling for spatial point processes.

2.1. *Modeling with spatial covariates.* We begin with a brief review of the basics of log Gaussian Cox process modeling. Consider a set of random locations which we denote by $S = \{\mathbf{s}_i\}_{i=1}^n$ where disease occurrence is observed over a spatial domain $D$. We model this random set of locations using a nonhomogeneous Poisson process with intensity function $\lambda(\mathbf{s})$ for all $\mathbf{s} \in D$. Let $\mathbf{z}(\mathbf{s})$ be a vector of location-specific covariates corresponding to a disease case observed at $\mathbf{s}$. For us, a key component of $\mathbf{z}(\mathbf{s})$ is the indicator of whether the case is in the metro area or not. However, in other contexts, we could envision information such as elevation, climate, exposure to pollutants, and so on to be relevant. We model $\lambda(\mathbf{s}) = r(\mathbf{s})\pi(\mathbf{s})$, where $r(\mathbf{s})$ is the population density surface at location $\mathbf{s}$. In practice, we may create such a surface using GIS tools and census data, or we may just work with areal unit population counts, letting $r(\mathbf{s}) = n(A)/|A|$ if $\mathbf{s} \in A$, where $n(A)$ is the number of persons residing in $A$ and $|A|$ is the area of $A$. The error introduced by this admittedly crude estimate may be mitigated somewhat by resorting to the non-Bayesian estimating equation alternatives discussed in Section 1.3. Specifically, in this framework one could model the two point processes separately [Waagepetersen (2007)] or jointly [by an extension of Guan (2006)].

Returning to our framework, $r(\mathbf{s})$ serves as an offset and $\pi(\mathbf{s})$ is interpreted as a population adjusted (or *relative*) intensity, which we model on the log scale as

$$\pi(\mathbf{s}) = \exp(\mathbf{z}(\mathbf{s})'\boldsymbol{\beta} + w(\mathbf{s})), \tag{1}$$

where $w(\mathbf{s})$ is a zero-centered stochastic process, and $\boldsymbol{\beta}$ is an unknown vector of regression coefficients. If $w(\mathbf{s})$ is taken to be a Gaussian process, then the original point process is called a *log Gaussian Cox process* [LGCP; Møller and Waagepetersen (2004), page 72]. The likelihood associated with $\boldsymbol{\beta}$ and $w_D = \{w(\mathbf{s}) : s \in D\}$ given $S$ takes the form

$$L(\boldsymbol{\beta}, w_D; S) \propto \exp\left(-\int_D r(\mathbf{s})\pi(s)\,d\mathbf{s}\right) \times \prod_{\mathbf{s}_i \in S} r(\mathbf{s}_i)\pi(\mathbf{s}_i). \tag{2}$$

Operating formally, a prior on $w_D$ along with a prior on $\boldsymbol{\beta}$ completes the Bayesian specification. Inference proceeds from the posterior which, again formally, is $p(\beta, w_D|S) \propto L(\boldsymbol{\beta}, w_D; S)p(\boldsymbol{\beta})p(w_D)$. Of course, the Gaussian process is only defined through its finite dimensional distributions so that, practically, this posterior is viewed in terms of a finite collection of locations. This motivates discrete approximation of the stochastic integral as we discuss below. One discrete approximation partitions $D$ into a collection of sets (say, $A_i, i = 1, 2, \ldots, m$) and creates a Poisson likelihood for the counts given $\lambda(A_i)$. That is, it models $\log \pi(A_i)$, thus precluding use of point level covariate information. Moreover, since $\pi(A_i) = \int_{A_i} \pi(s) \neq \exp(\int_{A_i} (\mathbf{z}(\mathbf{s})'\boldsymbol{\beta} + w(\mathbf{s})) d\mathbf{s})$, it is inappropriate to utilize the latter, simpler integration. Indeed, ignoring this inequality can introduce ecological fallacy issues; see, for example, Wakefield and Salway (2001) for a discussion.

We pursue an alternative discrete approximation which still enables us to work at the point level. Suppose we replace $\int_D \lambda(\mathbf{s})$ with some choice of numerical integration. For the moment, we allow analytic possibilities as well as Monte Carlo versions, since in either case, we will end up replacing $w_D$ with a finite set, say, $w^* = \{w(\mathbf{s}_j^*), j = 1, 2, \ldots, T\}$. Then we revise (2) to

(3) $\quad L(\beta, w^*, w(\mathbf{s}_1), \ldots, w(\mathbf{s}_n); S) p(w^*, w(\mathbf{s}_i), \ldots, w(\mathbf{s}_n)) p(\boldsymbol{\beta}).$

Now, we only need to work with an $(n+T)$-dimensional random variable to handle the $w$'s, hence, their prior is just an $(n+T)$-dimensional multivariate normal distribution. Note that, in (3), we will require that $\mathbf{z}(\mathbf{s})$ be available at each $\mathbf{t}_j$; that is, we require the component $\mathbf{z}(\mathbf{s})$ surfaces over $D$. These surfaces are not viewed as random and may be interpolated or tiled, according to the nature of the information for the particular spatial covariate; we assume only that we can assign a value of $\mathbf{z}$ for each $\mathbf{s} \in D$.

2.2. *Introducing nonspatial covariate information.* So far we have indicated how to incorporate covariates that are spatially referenced into the modeling. In our setting, we seek to introduce nonspatial covariates which we think of as being of two types (though the distinction will depend upon the application). One type of covariate provides the "marks" leading to a *marked point process* model. For us, this covariate is cancer type (colon vs. rectum), and we are interested in whether the two cancer intensity patterns differ.

The second type of covariate we view as an "auxiliary" variable that provides additional information associated with intensity. For us, age and cancer stage are examples of such covariates. Clearly patient age is associated with cancer intensity, but the strength of this association may differ across cancers. We wish to adjust intensity to reflect patient age, analogous to the age standardization used in aggregated areal data settings.

In general, we view these latter covariates as continuous[2] and introduce a second argument into the definition of the intensity, yielding a surface in $\mathbf{s}$ and $\mathbf{v}$ over the product space $D \times \mathcal{V}$ (i.e., geographic space by covariate space). We then generalize (1) to

(4) $\quad \pi(\mathbf{s}, \mathbf{v}) = \exp(\beta_0 + \mathbf{z}(\mathbf{s})'\boldsymbol{\beta} + \mathbf{v}'\boldsymbol{\alpha} + (\mathbf{v} \otimes \mathbf{z}(\mathbf{s}))'\boldsymbol{\gamma} + w(\mathbf{s})),$

---

[2]In the case of a discrete valued covariate, any integrals over $\mathbf{v}$ in our development are replaced by sums.

where the Kronecker product $\mathbf{v} \otimes \mathbf{z}(\mathbf{s})$ denotes the set of all the first order multiplicative interaction terms between $\mathbf{z}(\mathbf{s})$ and $\mathbf{v}$. When a particular interaction term is not of interest, the corresponding coefficient in $\boldsymbol{\gamma}$ is set to zero. This expression envisions a conceptual intensity value at each $(\mathbf{s}, \mathbf{v})$ combination. The interaction terms between spatial and nonspatial covariates provide the ability to adjust the spatial intensity by individual risk factors. If we fix $v$ in (4), we can view $\lambda(\mathbf{s}, \mathbf{v}) = r(\mathbf{s})\pi(\mathbf{s}, \mathbf{v})$ as a "conditional" intensity at level $\mathbf{v}$. If we *integrate* over $\mathbf{v}$ (see below), we obtain the (cumulative) marginal intensity $\lambda(\mathbf{s})$ associated with $\pi(\mathbf{s}, \mathbf{v})$.

Now, introducing marks $k = 1, 2, \ldots, K$, a general additive form for the log relative intensity is

$$(5) \qquad \log \pi_k(\mathbf{s}, \mathbf{v}) = \beta_{0k} + \mathbf{z}'(s)\boldsymbol{\beta}_k + \mathbf{v}'\boldsymbol{\alpha}_k + (\mathbf{v} \otimes \mathbf{z}(\mathbf{s}))'\boldsymbol{\gamma}_k + w_k(\mathbf{s}).$$

We can immediately interpret the terms on the right side of (5). The global mark effect is captured with the $\beta_{0k}$. Therefore, there is no intercept in $\mathbf{z}(\mathbf{s})$ and we have mark-varying coefficients for the spatially-referenced covariates, reflecting the possibility that these covariates can differentially affect the intensity surfaces of the marks. Similarly, we have mark-varying coefficients for the nuisance variables. We also have mark-varying coefficients for the interaction terms, reflecting possibly different effects of the nonspatial covariates over spatial domains. Finally, we allow the spatial random effects to vary with mark, that is, a different Gaussian process realization for each $k$. Dependence in the $w_k(\mathbf{s})$ surfaces may be expected (say, increased intensity at $\mathbf{s}$ for one marked outcome encourages increased intensity for another at that $\mathbf{s}$), suggesting the need for a *multivariate* Gaussian process over the $w_k$. Both separable and nonseparable forms for the associated cross-covariance function are conveniently specified through *coregionalization* [Gelfand et al. (2004); Banerjee, Carlin and Gelfand (2004), Sections 7.1 and 7.2].

Reduced models of (5) are immediately available, including, for example, $w_k(\mathbf{s}) = w(\mathbf{s})$, $\boldsymbol{\beta}_k = \boldsymbol{\beta}$, and $\boldsymbol{\alpha}_k = \boldsymbol{\alpha}$. Another interesting reduced model obtains by setting $\boldsymbol{\gamma}_k = 0$, leading to

$$(6) \qquad \log \pi_k(\mathbf{s}, \mathbf{v}) = \beta_{0k} + \mathbf{z}'(\mathbf{s})\boldsymbol{\beta}_k + \mathbf{v}'\boldsymbol{\alpha}_k + w_k(\mathbf{s}).$$

This separable form enables us to directly study the effect of the marks on spatial intensity. Specifically, the intensity associated with (5) is

$$(7) \qquad \lambda_k(s, v) = \exp(\beta_{0k} + \mathbf{v}'\boldsymbol{\alpha}_k) \times r(\mathbf{s})\exp(\mathbf{z}'(\mathbf{s})\boldsymbol{\beta}_k + w_k(\mathbf{s})).$$

We see a factorization into nonspatial nuisance and spatial covariate terms. Presuming the former is integrable over $\mathbf{v}$, the latter, up to a constant, is the "marginal spatial intensity."

Integration of $\lambda_k(\mathbf{s}, \mathbf{v})$, based upon (5), can be computed analytically in most cases. When $\mathbf{v}$ is categorical, the likelihood integral involves only integration over the spatial domain $D$. When $\mathbf{v}$ is continuous, simple algebra shows

$$\int_{\mathcal{V}} \lambda_k(\mathbf{s}, \mathbf{v}) \, dv \, d\mathbf{s} = r(\mathbf{s}) \exp(\beta_{0k} + \mathbf{z}(\mathbf{s})'\boldsymbol{\beta}_k + w_k(\mathbf{s}))$$
$$\times \int_{\mathcal{V}} \exp(\mathbf{v}'\boldsymbol{\alpha}_k + (\mathbf{v} \otimes \mathbf{z}(\mathbf{s}))'\boldsymbol{\gamma}_k) \, d\mathbf{v}.$$

Suppose, for instance, that there is only one component in $\mathbf{z}(\mathbf{s})$ and one component in $\mathbf{v}$ having range $(v_l, v_u)$. Provided $\alpha_k + z(\mathbf{s})\gamma_k \neq 0$, the marginal intensity $\lambda_k(\mathbf{s})$ is then

$$\int_{\mathcal{V}} \lambda_k(\mathbf{s}, v) \, dv \, d\mathbf{s}$$
$$= r(\mathbf{s}) \exp(\beta_{0k} + \beta_k z(\mathbf{s}) + w_k(\mathbf{s})) \times \int_{\mathcal{V}} \exp(v(\alpha_k + z(\mathbf{s})\gamma_k)) \, dv$$
$$= r(\mathbf{s}) \exp(\beta_{0k} + \beta_k z(\mathbf{s}) + w_k(\mathbf{s}))$$
$$\times \frac{1}{\alpha_k + z(\mathbf{s})\gamma_k}[\exp(v_u(\alpha_k + z(\mathbf{s})\gamma_k)) - \exp(v_l(\alpha_k + z(\mathbf{s})\gamma_k))].$$

Turning to the revised likelihood associated with (5), let $\{(\mathbf{s}_{ki}, \mathbf{v}_{ki}), i = 1, 2, \ldots, n_k\}$ be the locations and nuisance covariates associated with the $n_k$ points having mark $k$. The likelihood becomes

$$(8) \qquad \prod_k \exp\left(-\int_D \int_{\mathcal{V}} \lambda_k(\mathbf{s}, \mathbf{v}) \, d\mathbf{v} \, d\mathbf{s}\right) \times \prod_k \prod_{\mathbf{s}_{ki}, \mathbf{v}_{ki}} \lambda_k(\mathbf{s}_{ki}, \mathbf{v}_{ki}).$$

Using the calculations above, the double integral becomes

$$\int_D \int_{\mathcal{V}} \lambda_k(\mathbf{s}, v) \, dv \, d\mathbf{s}$$
$$= \int_D \biggl( r(\mathbf{s}) \exp(\beta_{0k} + \beta_k z(\mathbf{s}) + w_k(\mathbf{s}))$$
$$\times \frac{1}{\alpha_k + z(\mathbf{s})\gamma_k}[\exp(v_u(\alpha_k + z(\mathbf{s})\gamma_k)) - \exp(v_l(\alpha_k + z(\mathbf{s})\gamma_k))] \biggr) d\mathbf{s},$$

provided that the set $\{\mathbf{s} : \alpha_k + z(\mathbf{s})\gamma_k = 0\}$ has Lebesgue measure zero. Hence, the difficulty in the likelihood evaluation is the same as in (2) and can be treated in the manner described in conjunction with (3). In this regard, note that we bound the components of $\mathbf{v}$ in order to integrate explicitly over $\mathbf{v}$. We do not have a stochastic integration with regard to $\mathcal{V}$ as we have over $D$. Of course, sensitivity to the chosen bounds should be investigated.

Last, in the case of $k = 2$ marks, a common alternative model specification is logistic regression, which views the mark as the response given the locations and covariates. This is conditioning in the opposite order from our model, which views the locations and covariates as random given the marks. In our dataset it seems more natural to compare point patterns for the two different types of cancer, rather than view cancer type as some sort of binary "response" to covariate information.

**3. Results.** We now present the results of our analysis of the MCSS colon and rectum cancer data. Previous studies suggest that covariates related to a patient's socioeconomic status (SES) may be related to the patient's risk factors through its impact on diet, health care quality, or propensity to seek care. While our dataset lacks any individual-level SES measures, from census data we have several related tract-level variables: percentage of farm population, percentage of rural population, percentage of people with less than high school education, percentage of minority population, and poverty rate. A preliminary population-adjusted nonspatial Poisson regression analysis of our data on these covariates revealed only poverty rate and the metro indicator as significant predictors.

In our initial model, we consider two location-specific covariates: $z_1(\mathbf{s})$, the metro area indicator, and $z_2(\mathbf{s})$, the poverty rate in the census tract containing $\mathbf{s}$. We also employ two nonlocation-specific covariates: $v_1$, cancer stage [set to 1 if the cancer is diagnosed "late" (regional or distant stage) and 0 otherwise], and $v_2$, the patient's age at diagnosis. The population density $r(\mathbf{s})$ we use for standardization is available at 2000 census tract level, meaning that we assume population density is constant across any tract. The integral of the intensity is approximated by a Monte Carlo sum using a predictive process approximation [Banerjee et al. (2008)]; see the supplemental article by Liang, Carlin and Gelfand (2009) for full details.

The left and middle columns of Figure 2 show maps of the raw mean nonspatially varying covariates (age and proportion diagnosed late), while the right column maps a crude estimate of relative intensity for colon cancer (top row) and rectum cancer (bottom row). Notice these summaries are presented at tract level, even though we have exact (or nearly exact) spatial coordinates here. In the first two columns, tracts containing no cases are simply shaded according to the overall observed mean values for each disease, which are 69.9 and –64.8 for age and 0.618 and 0.555 for proportion diagnosed late for colon and rectum cancer, respectively. None of these four maps show strong spatial patterns, though we do see several areas with higher than average age, late diagnosis fraction, or both. The right column maps the logs of the numbers of cases divided by total number of residents in each tract. These crude maps of the tract-level log relative intensity (unadjusted for any spatial or nonspatial covariates) show somewhat stronger spatial patterns and higher overall rates of colon cancer. The rectum cancer map features an interesting collection of low outlying values in several outer-ring suburban census tracts.

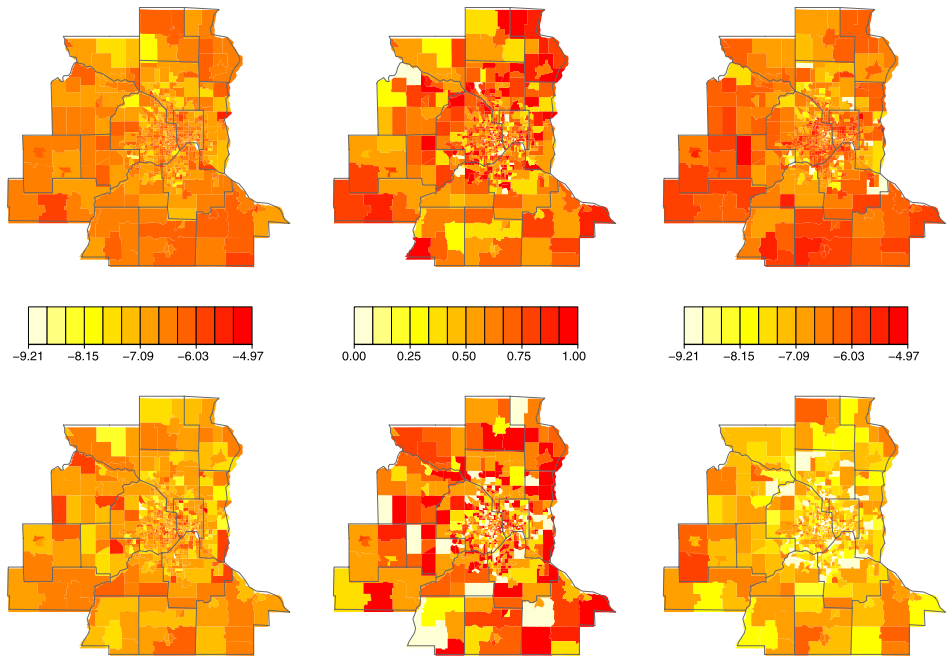

FIG. 2. *Minnesota colorectal cancer covariate and response data for colon (top row) and rectum (bottom row) groups: left, tract-specific map of observed mean age; middle, tract-specific map of observed proportion of late diagnosis; right, tract-specific observed log-relative intensity (count divided by population).*

TABLE 1
*Table of colorectum cancer patients' characteristics in metro and adjacent area of Minnesota. Percentages across appropriate columns are given in parentheses, and "ratio" gives the ratio of colon to rectum cases*

|  | **Total** | **Late = 0** | **Late = 1** | **Metro** | **Nonmetro** |
|---|---|---|---|---|---|
| All | 6544 | 2606 (40%) | 3938 (60.2%) | 5481 (83.8%) | 1063 (16.2%) |
| Colon | 4857 | 1855 (38%) | 3002 (61.8%) | 4079 (84%) | 778 (16%) |
| Rectum | 1687 | 751 (44.5%) | 751 (55.5%) | 1402 (83.1%) | 285 (16.9%) |
| Ratio | 2.88 | 2.47 | 4.0 | 2.91 | 2.73 |

Table 1 breaks down the data by stage and metro/nonmetro area. We see that 38% of colon cancer cases were diagnosed at an early stage, while 44.5% of rectum cancer cases were. In total, colon cancer is nearly three times as prevalent as rectum cancer in both the metro and nonmetro areas. A fact not revealed by the table is that there are 72 individuals who contribute *both* a colon and a rectum tumor. Since this is only around 1% of the total of 6544 individuals, we do not explicitly model this particular kind of dependence, but rather "lump it in" with the bivariate dependence modeled by $\rho$.

TABLE 2
*Model comparison using effective model size $p_D$ and DIC score. GLM refers to generalized linear model having no random effects*

| Model | $p_D$ | DIC |
|---|---|---|
| GLM (no residuals) | 11.8 | 1194.4 |
| Univariate spatial residuals | 72.0 | 692.4 |
| Bivariate spatial residuals | 80.2 | 688.8 |

Figure 3 shows tract-level maps of population density, $r(\mathbf{s})$, and our two location-specific covariates, $z_1(\mathbf{s})$ and $z_2(\mathbf{s})$. Not surprisingly, the central metro areas are the most populated. The poverty rate is fairly uniform except for high rates in a concentrated portion of the central metro.

We now fit our model, using independent Inverse Gamma$(2, 0.5)$ priors for $\sigma_1^2$ and $\sigma_2^2$, and a $Unif(-0.999, 0.999)$ prior for $\rho$. The scale of the spatial decay parameter $\phi$ is determined by the distance function employed. In this application, we started with a $Unif(130, 390)$ prior for $\phi$, so that the effective range lies between one-fourth and three-fourths of the maximal distance between any two knots. As expected, $\phi$ is only weakly identified, so a fairly informative prior is needed for satisfactory MCMC behavior. For simplicity, we simply fix the range parameter at $\phi = 195$, so that the effective range is roughly half of the maximal distance. A random-walk Metropolis–Hastings algorithm is used to draw posterior samples.

Table 2 compares the effective model size and DIC score of three models. It can be seen that the no-random effect model (GLM) is unacceptably bad, and the model with a single set of spatial residuals is not much worse than the bivariate residual model. This suggests that the two sets of residuals are fairly similar, and that $\rho$ is close to 1.

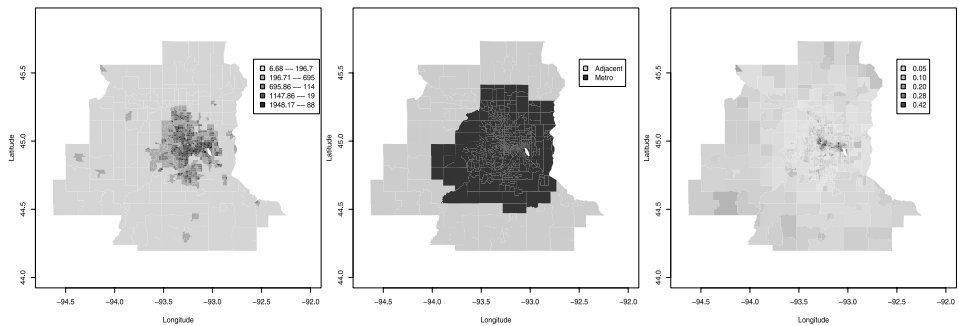

Fig. 3. *Left, population density by tract; middle, metro/nonmetro area; right, poverty rate by tract.*

Table 3 shows parameter estimates from some of our models. We parameterize so that the top rows concern the fixed effects for colon cancers, $\boldsymbol{\beta}_1$, but the second set of rows give the *differential* effect in the rectum cancer group, $\boldsymbol{\Delta} \equiv \boldsymbol{\beta}_2 - \boldsymbol{\beta}_1$. Thus, any 95% Bayesian confidence intervals that exclude 0 in this part of the table suggest a variable that has a significantly different impact on the two cancers.

In general, the effects of the non-spatial covariates are fairly similar across the models considered. We find that in the metro area there are relatively fewer cases of both colon and rectum cancer. This is consistent with statewide patterns of colorectal cancer occurrence in Minnesota, where higher age-adjusted rates are often found in nonmetro areas. However, there is no significant change in this relationship in the rectum group relative to the colon group. An interesting and somewhat counterintuitive finding is that poor areas seem to have relatively *fewer* cases. This appears consistent with the aforementioned finding of Wei et al. (2003) that colon cancer is associated with foods often consumed by relatively more affluent people (beef, pork, or lamb as a main dish, and other processed meat). However, unlike these authors, we find no significant difference in this relationship for rectum cancer.

TABLE 3
*Parameter estimates for the model with metro indicator and poverty rate as the spatial covariates, and stage and age as individual covariates. The estimates for rectum are relative effects to colon cancer. BSR = bivariate spatial residual model, USR = univariate spatial residual model, GLM = no random effects model*

|  |  | Fitted model | | |
|---|---|---|---|---|
|  |  | 93 knots | | |
|  |  | BSR | USR | GLM |
| Colon | intercept | $-8.76\,(-9.12,-8.44)$ | $-8.75\,(-9.25,-8.40)$ | $-8.91\,(-8.99,-8.83)$ |
|  | metro | $-0.23\,(-0.49,0.04)$ | $-0.19\,(-0.42,0.06)$ | $-0.21\,(-0.29,-0.14)$ |
|  | poverty | $-2.01\,(-2.47,-1.55)$ | $-1.90\,(-2.36,-1.47)$ | $-0.26\,(-0.61,0.09)$ |
|  | age | $0.36\,(0.31,0.40)$ | $0.36\,(0.31,0.40)$ | $0.32\,(0.28,0.36)$ |
|  | late | $0.48\,(0.42,0.54)$ | $0.48\,(0.42,0.54)$ | $0.48\,(0.43,0.54)$ |
|  | metro*age | $-0.06\,(-0.11,-0.02)$ | $-0.06\,(-0.11,-0.02)$ | $-0.06\,(-0.11,-0.02)$ |
| Rectum–colon | intercept | $-0.86\,(-1.08,-0.65)$ | $-0.84\,(-1.00,-0.68)$ | $-0.84\,(-1.01,-0.69)$ |
|  | metro | $0.02\,(-0.21,0.26)$ | $-0.07\,(-0.22,0.08)$ | $-0.07\,(-0.22,0.09)$ |
|  | poverty | $0.14\,(-0.70,0.98)$ | $-0.24\,(-1.06,0.52)$ | $-0.22\,(-1.00,0.49)$ |
|  | age | $-0.18\,(-0.26,-0.10)$ | $-0.18\,(-0.26,-0.10)$ | $-0.18\,(-0.25,-0.11)$ |
|  | late | $-0.26\,(-0.37,-0.15)$ | $-0.26\,(-0.38,-0.15)$ | $-0.26\,(-0.37,-0.15)$ |
|  | metro*age | $0.06\,(-0.03,0.15)$ | $0.05\,(-0.03,0.15)$ | $-0.01\,(-0.08,0.07)$ |
|  | $\rho$ | $0.98\,(0.95,0.99)$ | – | – |
|  | $\phi$ | 195 | 195 | – |
|  | $\sigma_1^2$ | $0.95\,(0.57,1.48)$ | $0.76(0.43,1.33)$ | – |
|  | $\sigma_2^2$ | $0.75\,(0.41,1.33)$ |  | – |

Turning to the nonlocation-specific covariates, age is significantly associated with increasing colon cancer, but a somewhat surprising relative *decrease* in rectum cancer. This difference ($-0.18$) is statistically significant, but not large enough in magnitude to make the overall age effect in the rectum group negative. A look at the data bears this out, with rectum cancers arising in a somewhat younger population; our preliminary Poisson regression also concurs, though here the relative decrease in the rectum group is not significant. Late detection provides another interesting difference between the colon and rectum groups: while there are significantly more cases diagnosed late than early, the effect of late diagnosis is significantly smaller in the rectum group (point estimate $-0.26$). Thus, public health interventions to encourage screening and early detection of colorectal cancer will have significantly greater impact on prevention for colon than for rectum. The metro-age interaction shows that the effect of age on colon cancer is significantly less pronounced in the metro area; a smaller "age adjustment" to the colon cancer intensity process is needed in the metro area. This effect is largely absent for rectum cancer, but this difference is not quite statistically significant. Finally, the estimate of $\rho$ is very close to 1, indicating very similar spatial residual patterns. This is perhaps a surprisingly strong association, but believable given that these are *residual* surfaces, which account (at least conceptually) for important missing covariates, which could be spatial (e.g., local screening percentage, other sociodemographic factors) or nonspatial (e.g., the physiological adjacency of the colon and the rectum).

Our Bayesian viewpoint allows us to make probabilistic statements both against and *in favor of* various null hypotheses of interest. For example, suppose we take 20% as the minimum difference in relative intensity required to conclude a practically meaningful difference between the colon and rectum cancer groups. For predictor $i$, this amounts to a test of $H_0 : \Delta_i \in [\log(0.8), \log(1.2)]$ versus $H_a : \Delta_i \notin [\log(0.8), \log(1.2)]$. The posteriors summarized in the second group of rows in Table 3 enable us to compute posterior odds ratios $OR = P(H_a|S)/P(H_0|S)$ for any predictor of interest. In our dataset, the two predictors of greatest substantive interest yield different results. Living in the metro area produces $OR = 0.12$, or odds of just over 8:1 in favor of no real difference in the colon and rectum groups. However, for late detection we obtain $OR = 2.81$, or nearly 3:1 odds in favor of a practically meaningful difference (in this case, a relative intensity reduction in the rectum group). Again, this suggests a public health program encouraging more aggressive cancer screening would be sensibly targeted to those living in regions with higher colon cancer relative intensity, since this should lead to a more meaningful reduction in cancer prevalence.

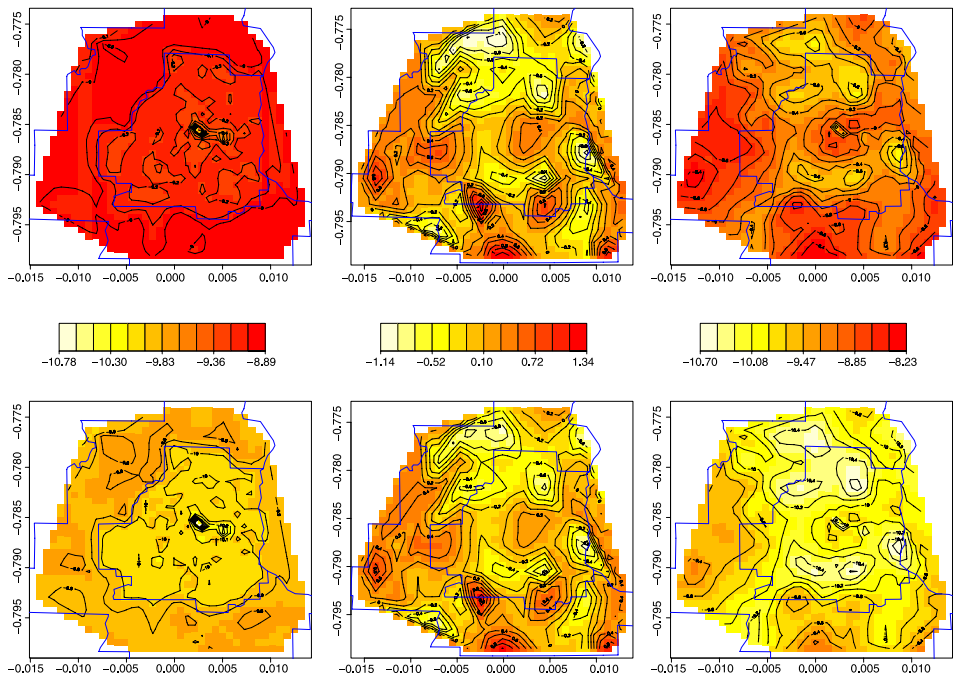

FIG. 4. *Log-relative intensity surfaces using values at centroid of each census tract at the mean age and assuming an early diagnosis. The top row is for colon cancer and the bottom for rectum cancer. The first column is the log-relative intensity surfaces without spatial residuals. The second column is the spatial residuals and the third is the complete log-relative intensity surfaces.*

Figure 4 shows maps of the fitted log intensity surfaces both without (left column) and with (right column) the spatial residuals (middle column), for a case at the mean age and diagnosed at an early stage. Without spatial residuals, the two spatial covariates alone predict slightly higher prevalence in the nonmetro areas. However, the residuals (which unlike our spatial covariates are point-level, and are thus summarized using image-contour maps) indicate further reductions are needed in the near southern, southeastern, and northern suburbs, as well as the far north exurban area. This leads to the more mottled fitted patterns in the rightmost column. Note these final two maps in the right column result in spatially smoothed versions of the corresponding maps in the right column of Figure 2, which we recall are something like raw log-relative intensity surfaces. While direct comparison is not really possible since the maps in the right column of Figure 4 *are* adjusted for both spatial and nonspatial covariates, the overall similarities further confirm the good fit scores achieved by our models. From a practical point of view, when combined with the significant differences in age and late detection between the two cancers found in Table 3, our findings encourage more aggressive colon cancer screening in the inner Twin Cities and the far southern and western exurbs, where the upper right panel of Figure 4 indicates higher colon cancer relative intensity.

**4. Discussion.** We have offered an analysis of colon and rectum cancer incidence data collected by the Minnesota Cancer Surveillance System during the period 1998–2002. In so doing we extended customary spatial point pattern analysis in the context of a log Gaussian Cox process model to accommodate covariates that are spatially referenced, individual-level cancer type marks, and individual-level risk factors that are not of interest in terms of marking. Our approach yields easy-to-interpret fixed effects for testing for equality of epidemiological properties across the two cancers, and fitted maps that can reflect the impacts of the spatially indexed covariates, spatial residuals, or both. These last maps also offer spatially smoothed fitted surfaces reminiscent of those in traditional areal models, but now adjusted for the nonspatially varying covariates, age and cancer stage.

As with many observational data analyses, our findings raise as many questions as they answer. The somewhat counterintuitive negative relationship between tract-level poverty and colorectal cancer shown in Table 3 might be the result of unmodeled confounding between age and poverty: poor areas could very well be significantly younger (especially in the metro, which features a higher proportion of immigrants, who tend to be younger). Since colorectal cancer is so highly associated with age, the apparent beneficial effect of poverty might just be another manifestation of the protective effect of youth. Similarly, modestly negative metro-age interaction may be due to more common use of colorectal screening in the metro area. Such screening methods can reduce colorectal cancer incidence by identifying pre-malignant lesions (polyps) and removing them; failure to screen a population might increase both the number of cases and the ages at which the cancers were diagnosed. Sadly, we currently lack the individual-level income and screening information necessary to precisely address these questions. Moreover, the MCSS database also does not feature information on diet, exercise, or family histories of the patients, the three previously-identified factors most likely to be responsible for any differences between colon and rectum cancer relative hazards. The data collected by MCSS is determined by legislation, and to expand it in any way requires a change in Minnesota state law, attempts at which the Minnesota Department of Health prefers to keep as rare as possible. As a result, future research regarding epidemiological properties of colon and rectum cancer should perhaps focus on obtaining approval and funding for a follow-up questionnaire mailed to all MCSS patients.

Even more fundamentally, an increasing number of authors view the debate over whether colon and rectum cancers have different etiologies as misplaced, arguing that the real distinction is not colon versus rectum, but rather proximal (right, or ascending) versus distal (left, or descending) colon, the latter of which includes the rectum. These authors argue that colorectal cancer is not a single disease, but two distinct diseases with distinct molecular profiles. One of these is more commonly found in the distal colon, and derives from hyperplastic polyps, whose putative successor lesions, serrated adenomas, represent discrete steps along a pathway to cancer [Huang et al. (2004)]. By contrast, the cancers most common in the proximal colon arise from an entirely different molecular pathway [O'Brien et al. (2006)]. Differences in risk factors for these two pathways are not well established but are nonetheless entirely likely. Relatedly, Glebov et al. (2003) found more than 1000 genes expressed differentially in adult ascending versus descending colon. Thus, the real subclassification of interest may not be colon versus rectum or distal colon versus proximal colon, but rather molecular pathology. Of course, this line of thinking encourages a reporting of cancers that is well beyond the capabilities of most U.S. public health reporting (hence intervention) systems, but the idea bears watching.

On the brighter side, recent audits have suggested our MCSS dataset is over 99% complete; that is, due to state reporting requirements, we are aware of essentially every tumor discovered by doctors in Minnesota. However, our methods obviously cannot reflect tumors that are not discovered or otherwise not reported. To the extent that such tumors happen unevenly across the spatial domain, this could lead to bias in our fitted estimates and maps. We do not think differential underreporting is a problem across our current, relatively compact and relatively urban 16-county spatial domain, but datasets that reached further into more remote regions of the state (especially semi-autonomous Native American tribal lands) may well suffer from this problem.

Future work in this area also includes extending our model with more complex interaction terms and perhaps more than two marks (the MCSS database has information on more than 20 cancers), leading to more challenging model-fitting. Another issue to address is the imprecision in the (typically rural) addresses within the point pattern. In some cases error may be simply due to the sensing device (e.g., the GPS unit), while in others it may be due to the practical limits of geocoding: for some of the cancers in our MCSS data, a significant proportion of the geocodes may be based on less than a complete and valid street address (e.g, residence zip + 2, residence zip only, or even the zip of a post office box). A final, perhaps most interesting path for the future lies in space–time point pattern analysis, in order to see evolution of cancer intensities over time. In the case of continuous time, we would now add a time argument to our intensity functions, leading to substantially increased scope for the modeling (e.g., separable versus nonseparable models for the space–time intensity). If time is instead viewed as discrete, we might instead extend our framework to temporally dynamic log

Gaussian Cox process models. Both of these options, while computationally challenging, could pay significant practical dividends.

**Acknowledgments.** The authors are grateful to Drs. Sudipto Banerjee, Andrew Flood, and Aaron Folsom for helpful discussions and Dr. Sally Bushhouse for permitting and facilitating our analysis of the Minnesota Cancer Surveillance System (MCSS) data.

## SUPPLEMENTARY MATERIAL

**Computational Issues** (DOI: 10.1214/09-AOAS240SUPP; .pdf). We provide full details of the Monte Carlo algorithms needed to approximate the complex point process likelihoods in the paper. In particular, we flesh out the details of our knot-based predictive process approximation, and give general guidelines for how the knots should be selected in any given application.

Shengde Liang
Bradley P. Carlin
Public Health in the Division of Biostatistics
School of Public Health
University of Minnesota
Minneapolis, Minnesota 55455
USA

Alan E. Gelfand
Department of Statistical Science
Duke University
Durham, North Carolina 27708
USA